\documentstyle[aps,prl,twocolumn,epsf,floats]{revtex}
\newcommand{\mrm}{\mathrm}

\begin{document}
\twocolumn[
  \begin{flushright}
  {\scriptsize \rm FERMILAB-PUB-99/022-T}\\
  {\scriptsize \rm UCSD/PTH/99-01}
  \end{flushright} 
  \vspace{-9mm}
  \begin{center}
  {\large\bf\ignorespaces
   Ab Initio Study of Hybrid ${\rm\overline{\mathbf{b}}\mathbf{gb}}$ Mesons}\\
   \bigskip   
   K.J.~Juge,$^1$ J.~Kuti$,^2$ and C.J.~Morningstar$^2$ \\
   {\small\it $^1$ Fermi National Accelerator Laboratory, P.O.~Box 500,
         Batavia, IL 60510 } \\ 
   {\small\it $^2$Dept.~of Physics, University of California at San Diego,
    La Jolla, California 92093-0319} \\ 
   \medskip
   {\small\rm (February 12, 1999)} \\
   \bigskip
  \begin{minipage}{5.5 true in} \small\quad
Hybrid ${\rm\overline{b}gb}$ molecules in which the heavy 
${\rm\overline{b}b}$ pair is bound together by the excited gluon field g
are studied using the Born-Oppenheimer expansion and 
quenched numerical simulations.
The consistency of results from the two approaches reveals a simple
and compelling physical picture for heavy hybrid states.
\medskip

 PACS number(s): 11.15.Ha, 12.38.Gc, 12.39.Mk
\end{minipage}
\end{center}
\vspace{4mm}
]

In addition to conventional hadrons, QCD predicts the existence of glueballs 
and hybrid states which contain excited gluon fields. 
Hybrid mesons with heavy ${\rm\overline{b}b}$ quark pairs
are the most amenable to theoretical
treatment. They are also experimentally accessible: early results from the
CUSB and CLEO collaborations\cite{cusb,cleo} revealed a complex resonance
structure between the $\rm{\overline{B}B}$ threshold and 11.2 GeV in 
$\rm{e^+e^-}$
annihiliation, precisely where the lowest hybrid excitations 
are expected\cite{kuti}.

In this work, we determine the masses of the 
lowest ${\rm\overline{b}gb}$ states.
Heavy hybrid mesons can be studied not only directly by numerical
simulation, but also using the Born-Oppenheimer expansion which is   
our primary guidance for the development of a simple
physical picture. The Born-Oppenheimer picture was introduced for the
description of heavy hybrid states in Refs.~\cite{hasenfratz,mandula} 
and was applied using hybrid potentials
first calculated in lattice QCD in Ref.~\cite{michael}.
In this new study, 
we work to leading order in the expansion and neglect higher-order
terms involving spin, relativistic, and retardation effects.  We test
the accuracy of the Born-Oppenheimer approach by comparison 
with high-precision results from simulations. 

\begin{figure}[t]
\begin{center}
\leavevmode
\epsfxsize=3.25in \epsfbox[167 354 488 582]{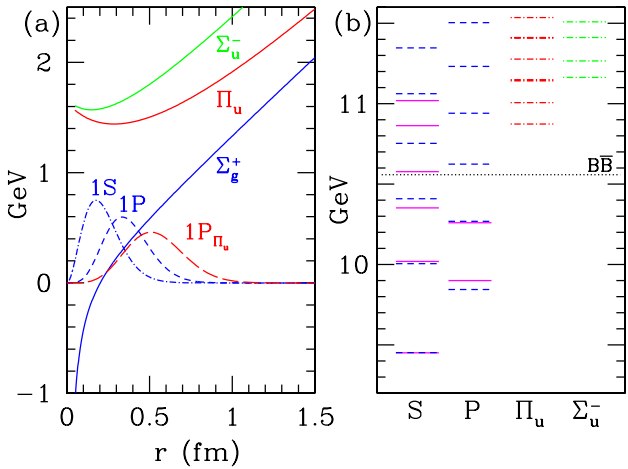}
\end{center}
\caption[figwf]{(a) Static potentials and radial probability densities
 against quark-antiquark separation ${\rm r}$ for 
 ${\rm r_0^{-1}=450}$ MeV.  The
 $\Sigma_{\mrm{g}}^+$ potential becomes the familiar running Coulomb law as
 ${\rm r}$ becomes very small. (b) Spin-averaged ${\rm \bar bb}$ spectrum
 in the LBO approximation (light quarks neglected).  Solid lines
 indicate experimental measurements.  Short dashed lines indicate 
 the $\mrm{S}$ and
 $\mrm{P}$ state masses obtained from the Schr\"odinger equation with the
 $\Sigma_{\mrm{g}}^+$ potential for ${\rm M_b=4.58}$ GeV. Dashed-dotted
 lines indicate the hybrid quarkonium states obtained 
 from the $\Pi_{\mrm {u}}$
 ${\rm (L=1,2,3)}$ and $\Sigma_{\mrm{u}}^-$ ${\rm (L=0,1,2)}$ potentials.
 Repeating the same analysis in the ${\rm\overline{c}gc}$ system,
 we find the lowest $\Pi_{\mrm{u}}$ hybrid ${\rm 1.19~GeV}$ above the spin
 averaged ground state.
\label{fig:LBO}}
\end{figure}

Our hybrid meson simulations are the first to
exploit anisotropic lattices with improved actions; 
preliminary reports on some of our results have
appeared previously\cite{earlier}. 
The hybrid meson mass uncertainties 
with improved anisotropic lattice technology are
dramatically smaller than those obtained in recent isotropic lattice
studies in the nonrelativistic formulation of lattice QCD (NRQCD)
using the Wilson gauge action\cite{sara,ukqcd}. 
We report here our final analysis on 
four distinct hybrid ${\rm\overline{b}gb}$ states. 
Although the effects of dynamical sea quarks are not included in
our quenched simulations, we will comment on their
impact on the hybrid spectrum.  The mass of the lowest hybrid 
${\rm\overline{c}gc}$ state was determined recently\cite{milc} without
NRQCD expansion for the slowly moving heavy quark and agrees
with our Born-Oppenheimer results\cite{kuti} 
(see caption of Fig.~\ref{fig:LBO}).
 
The hybrid meson can be treated analogous to a diatomic molecule:
the slow heavy quarks correspond to the nuclei and the fast gluon
field corresponds to the electrons\cite{hasenfratz}.
First, one treats the quark $\mrm{Q}$ and antiquark $\overline{\mrm{Q}}$ as
spatially-fixed color sources and determines the
energy levels of the excited gluon field 
as a function of the $\overline{\mrm{Q}}\mrm{Q}$ separation $\mrm{r}$; 
each of these excited energy levels defines an adiabatic potential
$V_{\overline{\mrm{Q}}\mrm{gQ}}(\mrm{r})$.
The quark motion is then restored
by solving the Schr\"odinger equation in each of these potentials.
Conventional quarkonia are based on the lowest-lying static potential;
hybrid quarkonium states emerge from the excited potentials.  Once the
static potentials have been determined (via lattice simulations),
it is a simple matter to determine the complete spectrum of
conventional and hybrid
quarkonium states in the leading Born-Oppenheimer (LBO) approximation.
This is a distinct advantage over meson simulations which yield only
the very lowest-lying states, often with large statistical uncertainties.
In addition, the LBO wave functions yield valuable
information concerning the structures
and sizes of these states which should greatly facilitate phenomenological
applications. 

The energy spectrum of the excited gluon field 
in the presence of a static quark-antiquark pair
has been determined in previous lattice
studies\cite{earlier}.  The three lowest-lying levels are shown
in Fig.~\ref{fig:LBO}.  These levels correspond to energy eigenstates
of the excited gluon field
characterized by the magnitude $\Lambda$ of the projection
of the total angular momentum ${\bf J}_{\mrm{g}}$ of the gluon field 
onto the molecular
axis, and by $\eta=\pm 1$, the symmetry quantum number
under the combined operations of
charge conjugation and spatial inversion about the midpoint between the quark
and antiquark of the ${\rm\overline{Q}gQ}$ system.  
Following notation from molecular spectroscopy,
states with $\Lambda=0,1,2,\dots$ are typically denoted
by the capital Greek letters $\Sigma, \Pi, \Delta, \dots$, respectively.
States which are even (odd) under the above-mentioned 
parity--charge-conjugation
operation are denoted by the subscripts $\mrm{g}$ ($\mrm{u}$).  
There is an additional label
for the $\Sigma$ states; $\Sigma$ states which are even (odd) 
under a reflection
in a plane containing the molecular axis are 
denoted by a superscript $+$ $(-)$.
In Ref.~\cite{earlier}, the potentials are calculated in terms of the hadronic
scale parameter ${\rm r_0}$ \cite{sommer}; the curves in Fig.~\ref{fig:LBO}
assume ${\rm r_0^{-1}=450}$ MeV (see below).  Note that as $\mrm{r}$ 
becomes small
(below 0.1 fm), the gaps between the excited levels and the 
$\Sigma_{\mrm{g}}^+$
ground state will eventually exceed the mass of the lightest glueball.  When
this happens, the excited levels will become unstable against glueball decay.

Given these static potentials, the LBO spectrum is easily obtained by
solving the radial Schr\"odinger equation with a centrifugal factor
${\rm \langle {\bf L}_{\overline{Q}Q}^2\rangle = L(L+1) - 2\Lambda^2
 + \langle {\bf J}_g^2 \rangle}$, where 
 ${\rm {\bf L}_{\overline{Q}Q}}$ is the orbital angular momentum
of the quark--antiquark pair.  For the $\Sigma_{\mrm{g}}^+$ potential,
$\langle {\bf J}_{\mrm{g}}^2 \rangle=0$. For the $\Pi_{\mrm{u}}$ and
$\Sigma_{\mrm{u}}^-$ levels, we attribute the lowest nonvanishing value
$\langle {\bf J}_{\mrm{g}}^2 \rangle=2$ to the excited gluon field.
Let ${\bf S}$ be the sum of the spins of the quark and antiquark, then
the total angular momentum of a meson is given by
${\bf J}={\bf L}+{\bf S}$.  In the LBO approximation, the eigenvalues
${\rm L(L+1)}$ and ${\rm S(S+1)}$ of ${\bf L}^2$ and ${\bf S}^2$ are 
good quantum
numbers.  The parity ${\rm P}$ and charge conjugation 
${\rm C}$ of each meson is given
in terms of ${\rm L}$ and ${\rm S}$ by 
${\rm P = \epsilon\ (-1)^{L+\Lambda+1}}$ and
${\rm C = \epsilon\ \eta\ (-1)^{L+\Lambda+S}}$, where ${\rm L\geq \Lambda}$ and
$\epsilon=1$ for $\Sigma^+$, $\epsilon=-1$ for $\Sigma^-$, and
$\epsilon=\pm 1$ for $\Lambda>0$.  Note that for each static potential,
the LBO energies depend only on ${\rm L}$ and the radial 
quantum number ${\rm n}$.

\epsfverbosetrue

Results for the LBO spectrum of conventional ${\rm \overline bb}$
and hybrid
${\rm\overline{b}gb}$ states are shown in~Fig.~\ref{fig:LBO}.
The heavy quark mass ${\rm M_b}$ is tuned to reproduce the 
experimentally-known
${\rm \Upsilon(1S)}$ mass: ${\rm M_\Upsilon=2M_b+E_0}$, 
where ${\rm E_0}$ is the energy of the
lowest-lying state in the $\Sigma_{\mrm{g}}^+$ potential. 
Level splittings are insensitive to small changes in the heavy quark mass.
For example, a $5\%$ change in ${\rm M_b}$ results in changes to the
splittings (with respect to the $1S$ state) ranging from $0.1-0.8\%$.

Below the $\overline{B}B$
threshold, the LBO results are in very good agreement with the
spin-averaged experimental measurements of bottomonium states.   
Above the
threshold, agreement with experiment is lost, suggesting significant
corrections from higher order effects and possible
mixings between the states from different adiabatic potentials.  
The mass of the lowest-lying hybrid (from the $\Pi_{\mrm{u}}$ potential) is
about 10.9~GeV.  Hybrid mesons from all other hybrid potentials are
significantly higher lying.  The radial probability 
densities for the conventional ${\rm 1S}$
and ${\rm 1P}$ states are compared with that of the lowest-lying
$\Pi_{\mrm{u}}$ hybrid state in Fig.~\ref{fig:LBO}.  Note that the 
size of the hybrid state
is large in comparison with the ${\rm 1S}$ and ${\rm 1P}$ states. 
For all of the
hybrid states studied here, the wave functions are strongly suppressed
near the origin so that the hybrid masses cannot be
affected noticeably by the small-${\rm r}$ instability of the 
excited-state potentials
from ${\rm\overline{b}gb\rightarrow \overline{b}b + glueball}$
decay.

The applicability of the Born-Oppenheimer approximation relies on the
smallness of retardation effects. The difference between
the leading Born-Oppenheimer Hamiltonian and the lowest order
NRQCD Hamiltonian is the ${\rm \vec{p}\cdot\vec{A}}$ coupling between
the quark color charge {\em in motion} and the gluon field. This
retardation effect, which is not included in the
LBO spectrum, can be tested by comparing the LBO 
mass splittings with those determined from meson simulations in NRQCD.

In order to obtain the masses of the
first few excited hybrid states in a given symmetry channel, we obtained
Monte Carlo estimates for a matrix of
hybrid meson correlation functions 
${\rm C_{ij}(t)=\langle 0\vert M_i(t)M_j^\dagger(0)
\vert 0\rangle}$ at two different lattice spacings.
Because the masses of the hybrid mesons are expected to
be rather high and the statistical fluctuations large, it is crucial to
use anisotropic lattices in which the temporal lattice spacing ${\rm a_t}$ is
much smaller than the spatial lattice spacing ${\rm a_s}$.  
Such lattices have already
been used to dramatically improve our knowledge of the Yang-Mills glueball 
spectrum\cite{peardon}.  In our simulations, the gluons are described
by the improved gauge-field action of Ref.\cite{peardon}.  The couplings
$\beta$, input aspect ratios $\xi$, and lattice sizes for each simulation
are listed in Table~\ref{table:simparams}.  
Following Ref.~\cite{peardon},
we set the mean temporal link ${\rm u_t=1}$ and obtain the mean spatial 
link ${\rm u_s}$ from the spatial plaquette.  
The values
for ${\rm r_0}$ in terms of ${\rm a_s}$ corresponding to each 
simulation were determined
in separate simulations.  Further details concerning the calculation
of ${\rm r_0/a_s}$ are given in Ref.~\cite{peardon}.  Note that we set the
aspect ratio using ${\rm a_s/a_t=\xi}$ in all of our calculations.  By
extracting the static-quark potential from Wilson loops in various
orientations on the lattice\cite{anisotropic}, we have verified that
radiative corrections to the anisotropy ${\rm a_s/a_t}$ are small.
The heavy quarks are treated within the NRQCD framework
\cite{nrqcd}, modified for an anisotropic lattice.
The NRQCD action includes only a covariant temporal derivative and the
leading kinetic energy operator (with two other operators to remove
${\rm O(a_t)}$ and ${\rm O(a_s^2)}$ errors); 
relativistic corrections depending
on spin, the chromoelectric ${\bf E}$ and chromomagnetic ${\bf B}$ fields,
and higher derivatives are not included. 

\begin{table}[t]
\caption[tabone]{Simulation parameters and results.  The second
  errors listed in the results in the bottom six rows
  are due to uncertainties in setting the
  heavy quark mass.
  \label{table:simparams}}
\begin{center}
\begin{tabular}{llll}
 $(\beta,\xi)$     &  $(3.0,3)$          &  $(2.6,3)$          \\[2mm]
 ${\rm u_s^4}$           &  $0.500$            &  0.451              \\
 lattice           &  $15^3\!\times\!45$ &  $10^3\!\times\!30$ \\
 $\#$ configs, sources  &  201, 16080    &  355, 17040          \\
 ${\rm r_0/a_s}$         &  4.130(24)          &  2.493(9)           \\
 ${\rm (\zeta,n_\zeta)}$ &  $(0.25,15)$        &  $(0.15,10)$        \\
 ${\rm a_sM_0}$          &  $2.56$             &  $3.90$             \\[2mm]
 ${\rm a_sM^S_{\rm kin}}$ & $5.03(2)$          &  $8.21(1)$           \\
 ${\rm r_0\;\delta(1P-1S)}$         &  $0.959(8)(3)$   &  $0.998(6)(3)$      \\
 ${\rm r_0\;\delta(2S-1S)}$         &  $1.303(11)(10)$ &  $1.252(8)(10)$     \\
 ${\rm r_0\;\delta(H_1-1S)}$        &  $3.287(53)(20)$ &  $3.338(54)(20)$    \\
 ${\rm r_0\;\delta(H_2-1S)}$        &  $3.37(13)(1)$   &  $3.443(47)(10)$    \\
 ${\rm r_0\;\delta(H_3-1S)}$        &  $4.018(55)(12)$ &  $4.034(76)(12)$    \\
 ${\rm r_0\;\delta(H_1^\prime-1S)}$ &  $4.204(67)(21)$ &  $4.229(62)(21)$ 
\end{tabular}
\end{center}
\end{table}

Our meson operators ${\rm M_i(t)}$ are constructed on a 
given time-slice as follows.
First, the spatial link variables are smeared using the algorithm
of Ref.~\cite{APEsmear} in which every spatial link ${\rm U_j(x)}$ on the
lattice is replaced by itself plus $\zeta$ times the sum of its four
neighboring spatial staples, projected back into SU(3); this procedure
is iterated ${\rm n_\zeta}$ times, and we denote the final smeared link
variables by ${\rm \tilde{U}_j(x)}$.  Next, let ${\rm \psi(x)}$ and 
${\rm \chi(x)}$ denote
the Pauli spinor fields which annihilate a heavy quark and antiquark,
respectively.  Note that the antiquark field is
defined such that ${\rm C\psi(x)C^\dagger=i\sigma_y\chi^\ast(x)}$, 
where ${\rm C}$
is the charge conjugation operator.  We define a {\em smeared}
quark field by
$
\tilde{\psi}(\mrm{x}) \equiv \left(1+\varrho\ a_{\mrm{s}}^2\tilde\Delta^{(2)}
 \right)^{n_\varrho}\psi(\mrm{x}),
$
 (and similarly for the antiquark field)
where $\varrho$ and ${\rm n_\varrho}$ are tunable parameters (we used
$\varrho=0.12,0.14$ and ${\rm n_\varrho=2-7}$) and the
covariant derivative operators are defined in terms of the smeared
link variables ${\rm \tilde{U}_j(x)}$.  These field operators, in addition
to the chromomagnetic field, are then used to construct our meson
operators, which are listed in Table~\ref{table:mesonops}.  The standard
clover-leaf definition of the chromomagnetic field $\tilde{\bf B}$
is used, defined also in terms of the smeared link variables.  Note that
four operators are used in each of the $0^{-+}$ and $1^{--}$
sectors. Because our NRQCD action includes
no spin interactions, we use only spin-singlet operators.  We can easily
couple these operators to the quark-antiquark spin to obtain various
spin-triplet operators, and the masses of such states will be degenerate
with those from the spin-singlet operators.

\begin{table}[t]
\renewcommand{\arraystretch}{1.2}  
\caption[tabtwo]{
  The meson spin-singlet operators used
  in each total angular momentum ${\rm J}$,
  parity ${\rm P}$, and charge conjugation ${\rm C}$ channel.  
  Note that ${\rm p=0,1,2,}$ and $3$
  were used to produce four distinct operators in the $0^{-+}$
  and $1^{--}$ sectors.  In the third column are listed the spin-triplet
  states which can be formed from the operators in the last column; the
  states in each row are degenerate for the NRQCD action used here.
 \label{table:mesonops}}
\begin{center}
\begin{tabular}{clcc}
 $J^{PC}$ & & Degeneracies & Operator \\ \hline 
 $0^{-+}$ & ${\rm S}$ wave & $1^{--}$ &$\tilde\chi^\dag\ \left[\tilde\Delta^{(2)}
     \right]^p\ \tilde\psi$ \\
 $1^{+-}$ & ${\rm P}$ wave & $0^{++},1^{++},2^{++}$ &$\tilde\chi^\dag
    \ \tilde{\mbox{\boldmath{$\Delta$}}} \ \tilde\psi$ \\
 $1^{--}$ & ${\rm H_1}$ hybrid & $0^{-+},1^{-+},2^{-+}$ &
    $\tilde\chi^\dag\ \tilde{\bf B}
       \left[\tilde\Delta^{(2)}\right]^p\ \tilde\psi$ \\
 $1^{++}$ & ${\rm H_2}$ hybrid & $0^{+-},1^{+-},2^{+-}$ & $\tilde\chi^\dag\ 
      \tilde{\bf B}\!\times\!\tilde{\mbox{\boldmath{$\Delta$}}}
      \ \tilde\psi$ \\
 $0^{++}$ & ${\rm H_3}$ hybrid & $1^{+-}$ & $\tilde\chi^\dag\ \tilde{\bf B}
     \!\cdot\!\tilde{\mbox{\boldmath{$\Delta$}}}\ \tilde\psi$ \\
\end{tabular}
\end{center}
\end{table}

In each simulation, the bare quark mass ${\rm a_sM_0}$ is 
set by matching the ratio 
${\rm R=M_{\rm kin}^S/\delta (1P-1S)}$, where ${\rm M_{\rm kin}^S}$ is 
the so-called kinetic mass
of the ${\rm 1S}$ state and ${\rm \delta (1P-1S)}$ is the energy separation 
between the ${\rm 1S}$
and ${\rm 1P}$ states, to its observed value $21.01(6)$.  The kinetic mass
${\rm M_{kin}^S}$ is determined by measuring the energy of the ${\rm 1S}$ state
for momenta ${\rm \vec{p}=(0,0,0)}$, ${\rm 2\pi(1,0,0)/L}$, and 
${\rm 2\pi(1,1,0)/L}$,
where ${\rm L}$ is the spatial extent of the periodic lattice.  These three
energies are then fit using ${\rm E_0+\vec{p}^2/(2M_{\rm kin}^S)}$ to extract
${\rm M_{\rm kin}^S}$.  Several low statistics runs using a range of quark
masses were done in order to tune the quark mass.  From the results of these
runs, we estimate that the uncertainty in tuning the quark mass is
about $5\%$.

\begin{figure}[t]
\begin{center}
\leavevmode
\epsfxsize=3.15in \epsfbox[28 144 572 508]{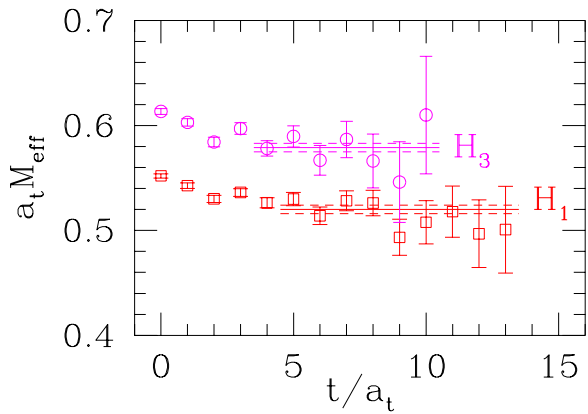}
\end{center}
\caption[figeffmass]{
  Effective masses ${\rm m_{\rm eff}(t)=-\ln[C(t+a_t)/C(t)]}$ 
  for two of the hybrids
  from the $(\beta,\xi)=(3.0,3)$ simulation.}
\label{fig:effmass}
\end{figure}

The simulation results are listed in Table~\ref{table:simparams}.
The masses ${\rm m_i}$ in the $1^{+-}$, $1^{++}$, and $0^{++}$ channels 
are extracted
by fitting the single correlators ${\rm C_i(t)}$ to their expected 
asymptotic form
${\rm C_i(t)\rightarrow Z_i\exp(-m_i t)}$ for sufficiently large ${\rm t}$.  
In each of the
$0^{-+}$ and $1^{--}$ channels, we obtain a $4\times 4$ correlation
matrix.  The variational method is then applied to reduce the $4\times 4$
matrix down to an optimized $2\times 2$ correlation matrix 
${\rm C^{opt}_{ij}(t)}$.  For sufficiently
large ${\rm t}$, we fit all elements of this matrix using 
${\rm \sum_{p=0}^1 Z_{ip}Z_{jp}\exp(-m_p t)}$ 
to extract the two lowest-lying masses.
In this way, we obtain an estimate of the ${\rm 2S}$ mass, as well as the
first excited hybrid state ${\rm H_1^\prime}$.  
The effective masses corresponding
to several of the correlation functions obtained in the $\beta=3.0$, $\xi=3$
simulation are shown in Fig.~\ref{fig:effmass}.

\begin{figure}[t]
\begin{center}
\leavevmode
\epsfxsize=3.15in \epsfbox[28 144 572 552]{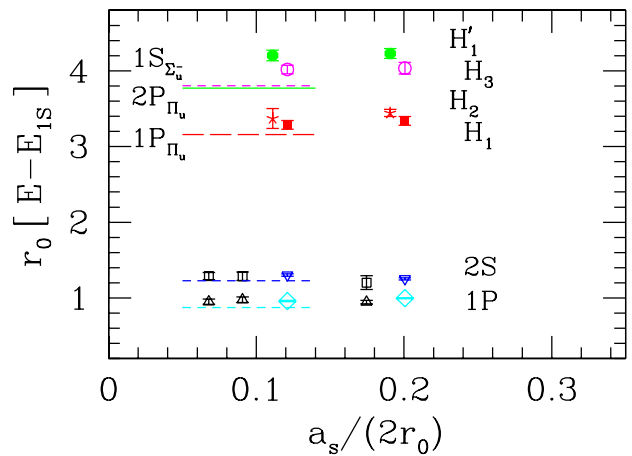}
\end{center}
\caption[figscaling]{
  Simulation results for the level splittings (in terms of ${\rm r_0}$ and
  with respect to the $1S$ state) against the lattice spacing ${\rm a_s}$.  
  Results
  from Ref.~\protect\cite{Wilson} using an NRQCD action with higher-order
  corrections are shown as $\Box$ and $\bigtriangleup$; 
  all other
  symbols indicate results from this work.  Some points have been shifted
  horizontally to prevent overlaps.  The horizontal lines show the
  LBO predictions.}
\label{fig:scaling}
\end{figure}

The simulation results for the level splittings (in terms of ${\rm r_0}$ 
and with
respect to the ${\rm 1S}$ state) are shown in Fig.~\ref{fig:scaling} 
against the
lattice spacing.  Small finite--${\rm a_s}$ errors are evident in the
${\rm 1P}$ and ${\rm 2S}$ splittings from the coarse lattice simulation; 
none of the
four hybrid splittings show any significant discretization errors.  The
simulation results compare remarkably well with the LBO predictions, shown
as horizontal lines in Fig.~\ref{fig:scaling}.  In the LBO approximation,
the ${\rm H_1}$ and ${\rm H_2}$ mesons correspond to degenerate 
${\rm 1P_{\Pi_u}}$ states
of opposite $\epsilon$, the ${\rm H_3}$ hybrid corresponds to a 
${\rm 1S_{\Sigma_u^-}}$
state, and the ${\rm H_1^\prime}$ corresponds to a 
${\rm 2P_{\Pi_u}}$ level; furthermore,
the ${\rm H_3}$ and ${\rm H_1^\prime}$ hybrids are predicted to be 
nearly degenerate,
with the ${\rm H_1^\prime}$ lying slightly lower.  
The simulation results share
these same qualitative features, except that the ${\rm H_1^\prime}$ 
lies slightly
higher than the ${\rm H_3}$.  The LBO approximation
reproduces all of the level splittings to within $10\%$.  In 
Fig.~\ref{fig:scaling}, we also show results\cite{Wilson} for the ${\rm 1P}$
and ${\rm 2S}$ splittings for an NRQCD action including higher order
relativistic and spin interactions; the effects of such terms are seen
to be very small.  Note that
spin-dependent mass splittings are difficult to estimate 
in hybrid states since the excited gluon field extends
on the scale of the confinement radius with a nonperturbative 
wavefunction when its color magnetic moment interacts with 
the heavy-quark spins.

To convert our mass splittings into physical units, we must specify the
value of ${\rm r_0}$.  Using the observed value for the 
${\rm 1P-1S}$ splitting, we
find that ${\rm r_0^{-1} = 467(4)}$ MeV; using the ${\rm 2S-1S}$ 
splitting, we
obtain ${\rm r_0^{-1} = 435(5)}$ MeV.  This discrepancy is caused by our
neglect of light quark effects\cite{quenching}.  
Taking ${\rm r_0^{-1}=450(15)}$
MeV, our lowest-lying hybrid state lies $1.49(2)(5)$ GeV (the second error
is the uncertainty from ${\rm r_0}$) above the ${\rm 1S}$ state.

Hybrid and conventional states substantially extending over 1 fm
in diameter are vulnerable to light-quark vacuum polarization
loops which will dramatically change the static potentials
through configuration mixing with ${\rm\overline B B}$ mesons; instead of
rising indefinitely with ${\mrm r}$, these potentials will eventually
level off since the heavy ${\rm \overline{\mrm{Q}}\mrm{gQ}}$ state can
undergo fission into two separate ${\rm\overline{Q}q}$ color singlets,
where ${\rm q}$ is a light quark.  We expect that the plethora of hybrid
and conventional states above 11 GeV obtained from the quenched potentials
will not survive this splitting mechanism as observable resonances.
For quark-antiquark separations below 1.2 fm or so, there is evidence from
recent studies\cite{bali,kuti} that light-quark vacuum polarization effects
do not appreciably alter the $\Sigma_{\mrm{g}}^+$ and $\Pi_{\mrm{u}}$
inter-quark potentials (for light-quark masses such that 
${\rm m_\pi \sim m_\rho/2}$).  Since such distances are the most relevant
for forming the lowest-lying bound states, the survival of the lightest
${\rm\overline{b}gb}$ hybrids as well-defined resonances above the
${\rm\overline B B}$ threshold remains conceivable.

During the preparation of this work we learned about new results \cite{CPpacs}
which have considerable overlap with our NRQCD simulations. 

This work was supported by the U.S.~DOE,
Grant No.\ DE-FG03-97ER40546.


\begin{references}
\vspace{-13mm}
\bibitem{cusb}
   D.~Lovelock {\it et al.}, Phys.\ Rev.\ Lett.\ {54}, 377 (1985).
\bibitem{cleo}
   D.~Besson {\it et al.}, Phys.\ Rev.\ Lett.\ {54}, 381 (1985).
\bibitem{kuti}
   J.~Kuti, Proceedings of the XVI International Symposium on Lattice 
   Field Theory, Nucl.\ Phys.\ B(Proc.\ Suppl.) in press; hep-lat/9811021. 
\bibitem{hasenfratz}
   P.~Hasenfratz, R.~Horgan, J.~Kuti, J.~Richard, 
   Phys.\ Lett.\ B{95}, 299 (1980). 
\bibitem{mandula} 
    D.~Horn and J.~Mandula,  
    Phys.\ Rev.\ D{17}, 898 (1978).   
\bibitem{michael}
   S.~Perantonis and C.~Michael, Nucl.~Phys., B347 (1990) 854.    
\bibitem{earlier}
   K.J.~Juge, J.~Kuti, and C.~Morningstar, Nucl.\ Phys.\ B(Proc.\ Suppl.) {63} 
   326, (1998); and unpublished (hep-lat/9809098). 
\bibitem{sara}
  S.~Collins, C.~Davies, G.~Bali, Nucl.\ Phys.\ B(Proc.\ Suppl.){63},
  335 (1998). 
\bibitem{ukqcd}
  T.~Manke {\it et al.}, Phys.\ Rev.\ D{57}, 3829 (1998).  
\bibitem{milc}
  C.~Bernard {\it et al.,} Phys.\ Rev.\ D{56}, 7039 (1997).              
\bibitem{sommer}
   R.~Sommer, Nucl.\ Phys.\ {B411}, 839 (1994).   
\bibitem{peardon}
   C.~Morningstar and M.~Peardon, Phys.\ Rev.\ D{56}, 4043 (1997);
   and to appear (hep-lat/9901004).  
\bibitem{anisotropic}
  C.~Morningstar, Nucl.\ Phys.\ B(Proc.\ Suppl.){53}, 914 (1997).    
\bibitem{nrqcd}
   G.~P.~Lepage, {\it et al.,}
   Phys.\ Rev.\ D{46}, 4052 (1992).   
\bibitem{APEsmear}
   M.~Albanese {\it et al.,} Phys.\ Lett.\ B{192}, 163 (1987).
\bibitem{Wilson}
  C.~Davies {\it et al.}, Phys.\ Rev.\ D{58}, 054505 (1998).
\bibitem{quenching}
  C.~Davies {\it et al.,} Nucl.\ Phys.\ B(Proc.\ Suppl.){47}, 409
 (1996); A.~Spitz {\it et al.}, Nucl.\ Phys.\ B(Proc.\ Suppl.){63},
  317 (1998).
\bibitem{bali}
   G.~Bali, private communication, unpublished.   
\bibitem{CPpacs}
  T.~Manke {\it et al.}, unpublished (hep-lat/9812017).  
\end{references}
\end{document}